\documentclass[twocolumn,preprintnumbers,amsmath,amssymb]{revtex4}
\usepackage{epsfig}
\usepackage{amsmath}
\usepackage[usenames,dvipsnames]{color}

\newcommand{\edu}[1]{{\color{black} #1}}

\usepackage[normalem]{ulem}

\begin{document}
\title{Homogeneous ice nucleation at moderate supercooling from molecular simulation}
\author{E. Sanz, C. Vega, J. R. Espinosa, R. Caballero-Bernal, J. L. F. Abascal and C. Valeriani}
\affiliation{Departamento de Qu\'{\i}mica F\'{\i}sica,
Facultad de Ciencias Qu\'{\i}micas, Universidad Complutense de Madrid,
28040 Madrid, Spain}
\date{\today}
\begin{abstract}
Among all the freezing transitions, that of water into ice is probably the most
relevant to biology, physics, geology or atmospheric science.
In this work we investigate homogeneous ice nucleation by means of
computer simulations. We evaluate the size of the critical cluster and the
nucleation rate for temperatures ranging between 15~K and 35~K below melting. 
We use the TIP4P/2005 and the TIP4P/Ice water models. Both 
give similar results when compared at the same temperature difference
with the model's melting temperature.  
The size of the critical cluster varies from $\sim$8000~molecules (radius$ =
4$~nm) at 15~K below melting to $\sim$600 molecules (radius$ = 1.7$~nm) at
35~K below melting. 
We use Classical Nucleation Theory (CNT) to estimate  
the ice-water interfacial free energy and the nucleation free energy barrier. 
We obtain an interfacial free energy of 29(3)~mN/m from an extrapolation of our results
to the melting temperature. 
This value is in good agreement both with experimental measurements 
and with previous estimates from computer simulations of TIP4P-like models.
Moreover, we obtain estimates of the nucleation rate from simulations of the 
critical cluster at the barrier top. The values we get for both models
agree within statistical error with experimental measurements. 
At temperatures higher than 20~K below melting we get nucleation rates slower than the
appearance of a critical cluster in all the water of the hydrosphere in the age
of the universe. Therefore, our simulations predict that water freezing above
this temperature must necessarily be heterogeneous.
\end{abstract}

\maketitle

\twocolumngrid

\section{Introduction}

When a liquid is cooled below its freezing point it is supposed to freeze.
Usually, impurities or the solid boundaries of the liquid provide preferential
sites for the formation of the solid phase.
However, even in the absence of impurities, small nuclei of the new phase may
be formed within the bulk metastable liquid.
This mechanism of formation of the solid phase is called homogeneous
nucleation.\cite{debenedettibook,kashchievbook}
Homogeneous nucleation is an activated process since the formation of a
critical nucleus requires the surmounting of a free energy barrier.
After that, the crystalline nucleus can grow (nucleation-and-growth mechanism).
In general, at moderate supercooling, the limiting step is the formation of
the critical cluster rather than the crystal growth. The most relevant
quantity to characterize nucleation is the nucleation rate, {\em i.e.} the
number of nucleating clusters per unit time and volume.

Water freezing is arguably the most important liquid-to-solid transition. 
For example, ice formation in atmospheric clouds is a key factor to the global radiation 
budget and to  
climate change.\cite{review_ice_formation_clouds_2005,baker97,demott10}
Water freezing is also a big issue, for instance, in the cryopreservation of cells and tissues.\cite{cryopres}
Moreover, ice formation is relevant to microbiology\cite{hirano00},
food industry\cite{maki74,bacterial_ice_nucleation}, materials 
science\cite{michaelides07}, geology\cite{weathering_book} 
or physics.\cite{pruppacher1995,debenedettibook,taborek,knopf11,nature_valeria,galli_mw,riechers13}

Despite its great importance, our understanding of water freezing is far from 
complete. Not even homogeneous nucleation, the simplest 
conceivable mechanism by which ice can be formed, is fully understood.
One of the reasons for this is the need to perform experiments 
with small droplets (10-100$~\mu$m) 
to avoid heterogeneous nucleation.\cite{kramer1999,stockel2005,murray2010,mishima10}
This, and the time that the droplets can be stabilized, sets the order of magnitude
that can be probed for the nucleation rate, $J$. Thus, experimental measurements for 
$\log_{10} (J /$(m$^{-3}$s$^{-1}$)) typically 
range between $4$ and and $14$ . This corresponds to a temperature window spanning from 239~K to 233~K, 
the latter often referred to as ``homogeneous nucleation temperature''.\cite{debenedetti03} 
Our knowledge of the nucleation rate outside this temperature window is limited
to extrapolations based on CNT.  Such extrapolations
must be taken with care since the uncertainties in the nucleation rate and the
narrow range of temperatures for which J can be measured lead to important
differences in the estimated value of the interfacial free energy and/or the
kinetic prefactor.\cite{murray2010}
Moreover, it has not been possible so far to observe a 
critical ice nucleus in experiments because critical nuclei 
are relatively small and short-lived. Therefore, we only have 
estimates of the critical cluster size based on experimental measurements
of $J$.\cite{pruppacher1995,furukawa1996,jeffery1997,kramer1999,kiselev01,micelas}
The purpose of this paper is to fill these gaps by obtaining
the first estimate of the size of the critical
cluster and of the nucleation rate at high temperatures which 
is not entirely based on theoretical extrapolations from measurements
at low temperatures. We will make use of computer simulations
to achieve these goals.

Computer simulations are a valuable tool to investigate nucleation\cite{S_1997_277_01975,Nature_2001_409_1020} since they
provide a microscopic description of the process.
It is therefore somehow surprising that the number of simulation studies dealing
with ice nucleation is rather small.\cite{sear2012} On the one hand, it has
been shown that ice nucleation can occur spontaneously (without the aid of 
special simulation techniques) when 
an electric field is applied\cite{kusalik_electric_field}, when  crystallization is assisted by a 
substrate\cite{koga01,moore10} or by an interface\cite{subsurface}, when coarse-grained models with accelerated dynamics are 
simulated at high supercoolings,\cite{nature_valeria,valeria_jcp_2010,valeria_pccp_2011} or when small systems 
are simulated\cite{matsumoto02,PhysRevLett.88.195701,st2_recent}. 
On the other hand, if nucleation does not
happen spontaneously, rare event techniques must be used. 
The number of such works is limited and the agreement
between different groups is not entirely satisfactory.
Radhakrishnan and
Trout\cite{trout03,trout_prl_2003}, Quigley and Rodger\cite{quigley08} and Brukhno et
al.\cite{anwar_tip4p_nucleation}  determined
the free energy barrier for the formation of ice critical clusters with the
TIP4P water model at 180~K (50 degrees below
the model's melting temperature), but mutually consistent results were not found.
Reinhardt and Doye\cite{doye12} and Li {\em et al.}\cite{galli_mw} evaluated
the nucleation rate of the mW model at 55 K below freezing finding a 
discrepancy of 
six orders of magnitude. 
Very recently, 
Reinhardt {\em et al.} investigated ice nucleation at moderate supercoolings,\cite{doyejcp2013} 
to estimate the free energy of formation of small pre-critical clusters.  
It is almost certain that more ice nucleation studies are on the way and,
hopefully, the discrepancies will become smaller.

None of the studies mentioned in the previous paragraph deal with large systems at moderate 
supercoolings like the present investigation does. 
By supercooling, $\Delta T$, we mean 
the difference between the melting temperature and the temperature of interest. 
Note that the melting temperature of a model does not necessarily coincide with 
the experimental melting temperature or with the melting temperature of other models.
In this work we determine, by means of
computer simulations, the size of critical ice clusters and the nucleation rate 
for $\Delta T$ ranging from 15 to 35~K. 
\edu{In this way we provide, for the first time, nucleation rates for $\Delta T$ lower
than 35 K, where experimental measurements  are not currently feasible (CNT based
estimates of $J$ can in principle be made for any supercooling but, to the best of our knowledge, 
there are no such estimates available  for $\Delta T < 30 K$).\cite{pruppacher1995,kashchievbook,murray2010}}
Our simulations predict that for $\Delta T < 20$~K
it is impossible that homogeneous ice nucleation takes place.
Therefore, ice must necessarily nucleate heterogeneously 
for supercoolings lower than 20~K.
Moreover, we can directly compare our results for the 
largest studied supercoolings to the experimental measurements. 
We find, within uncertainty, a good agreement 
with experimental nucleation rates. 
We 
predict that the radius of the critical cluster goes from $\sim$40~\AA (8000 molecules)
at $\Delta T$ {\em ca.} 15~K to $\sim$17~\AA (600 molecules)
at $\Delta T$ {\em ca.} 35~K.   
We also estimate the surface free energy via CNT. 
We obtain, in agreement with 
predictions based on experimental measurements,\cite{pruppacher1995,zobrist07,alpert11} that the surface free energy decreases 
with temperature. An extrapolation of the interfacial free energy to the melting
temperature gives a value of $\sim$29~mN/m, in reasonable 
agreement with experimental results\cite{gamma_exp}, and with
calculations by simulation.\cite{gammadavid}

We use two simple, yet realistic, water models; namely 
TIP4P/2005\cite{TIP4P2005} and TIP4P/Ice\cite{TIP4PICE}. The 
melting temperature\cite{TIP4P2005,TIP4PICE} and the ability of these models to 
predict properties of real water has already been well established.\cite{vega11}
The results obtained for both water models are quite similar provided that they are compared at the
same $\Delta T$.

\section{Methodology}

To evaluate the size of critical ice clusters we follow a similar approach
to that proposed by Bai and Li\cite{bai06} to calculate the solid-liquid
interfacial energy for a Lennard-Jones system. They employ spherical crystal
nuclei embedded in the supercooled liquid and determine the temperature at
which the solid neither grows nor melts.
The key issue of this methodology is that determining the melting temperature
of a solid cluster embedded in its corresponding supercooled liquid water is
equivalent to the determination of the critical size of the cluster for a
certain given temperature.
Thus, in a sense, this methodology can be regarded as the extension to nucleation
phenomena of the well known direct coexistence technique.\cite{woodcook2}
A similar method was applied to water by Pereyra et al.\cite{carignano} They
inserted an infinitely long (through periodical boundary conditions) ice
cylinder in water and determined the melting temperature of the
cylinder.
Recently, the approach of Bai and Li has been used to investigate  
the nucleation of clathrate 
hydrates.\cite{jacobson11,knott12}

Here we shall implement this methodology to study a three-dimensional spherical
ice cluster embedded in supercooled water. This follows closely the
experimental situation where the incipient ice embryo is fully immersed into
liquid water.
Such {\it brute force} approach requires very large systems
(containing up to 2$\times 10^5$ water molecules). 
However, molecular dynamics simulations can be efficiently parallelised so that it is 
nowadays possible to deal with such system size.
The methodology can then be implemented in a rather straightforward way, and
is particularly useful at moderate supercooling,
where other techniques (such as umbrella sampling\cite{ARPC_2004_55_333,auer_frenkel}, Forward Flux Sampling\cite{PRL_2005_94_018104} or 
Transition Path Sampling\cite{ARPC_2002_53_0291}) may become numerically too expensive.

Once we calculate the critical cluster size we make use of 
CNT \cite{ZPC_1926_119_277_nolotengo,becker-doring,kelton} in its version
for spherical clusters to 
estimate the surface free energy, $\gamma$:
\begin{equation}
\label{ncrit}
\gamma=\left(\frac{3 N_{c} \rho^2_s |\Delta \mu|^3}{32 \pi}\right)^{1/3}
\end{equation}
where  $\rho_s$ is the number density of the solid and $\Delta \mu$ is 
the chemical potential difference between the metastable liquid and the solid at the temperature under consideration.
This expression allows us to obtain a value for $\gamma$ associated to each cluster.
CNT can also be used to estimate the height of the nucleation free-energy barrier,  $\Delta G_c$: 
\begin{equation}
\label{eq_G_cnt}
\Delta G_c  = \frac{16 \pi \gamma^3}{3 \rho^2_s |\Delta \mu|^2}.
\end{equation}
Once $\Delta G_c$ is known, we can use the following CNT-based expression to evaluate the 
nucleation rate\cite{auerjcp}:
\begin{equation}
J=Z f^{+} \rho_f \exp(-\Delta G_{c}/(k_B T))
\label{eqrate}
\end{equation}
where $Z$ is the Zeldovich factor, $Z=\sqrt{(|\Delta G^{''}|_{N_c}/(2\pi k_B T))}$,
and $f^{+}$ is the attachment rate of particles to the critical cluster.
The CNT form of the Zeldovich factor is 
\begin{equation}
Z=\sqrt{|\Delta \mu|/(6 \pi k_B T N_c)},
\end{equation}
which can be obtained from our calculations of $N_c$.
We follow Ref.~\onlinecite{auerjcp} to calculate $f^{+}$ as a diffusion coefficient of the 
cluster at the top of the barrier: 
\begin{equation}
f^{+}=\frac{<(N(t)-N_c)^2>}{2t}. 
\label{eqattach}
\end{equation}
Therefore, in order to obtain nucleation rates we combine CNT predictions with simulations
of the critical clusters. 

By using the methodology here described, the nucleation rate of clathrate hydrates 
has been recently calculated.\cite{knott12} The validity of this approach 
relies on the ability of CNT to make good estimates of the free energy barrier
from measured values of the critical cluster size. CNT is expected to work well for big 
critical clusters. We are confident that the cluster sizes we deal with in this work 
are big enough for CNT to produce meaningful predictions. 
We discuss why in Sec.~\ref{validity}.

\section{Technical details}

\subsection{Simulation details}
We carry out $NpT$ GROMACS\cite{hess08}  molecular dynamics simulations (MD)  
of a system that consists of one spherical ice-Ih cluster surrounded by  supercooled water  molecules.  
We use two different rigid non-polarizable models of water:   
TIP4P/2005\cite{TIP4P2005} and TIP4P/Ice.\cite{TIP4PICE}
TIP4P/2005 is a  model that provides a quantitative account of many water
properties\cite{vega09,vega11} including not only the well known thermodynamic
anomalies but also the dynamical ones.\cite{pi09,gonzalez10}
TIP4P/Ice was designed to reproduce the melting temperature, the densities and
the coexistence curves of several ice phases. 
One of the main differences between the two models is 
their ice Ih melting temperature at 1~bar: $T_m = 252$~K for TIP4P/2005 and
$T_m = 272$~K for TIP4P/Ice.
We evaluate long range  electrostatic interactions using the smooth
Particle Mesh Ewald method\cite{essmann95} and  truncate both the LJ and real
part of the Coulombic interactions at 9~\AA. 
We preserve the rigid geometry of the water model  by
using constraints. All simulations are run at the constant pressure of
$p = 1$~bar, using an isotropic Parrinello-Rahman barostat\cite{parrinello81}
and  at constant temperature, using the velocity-rescaling
thermostat.\cite{bussi07}
We set the MD time-step to  3~fs.

\subsection{Order parameter}
\begin{figure}[h!]
\includegraphics[width=0.4\textwidth,clip=]{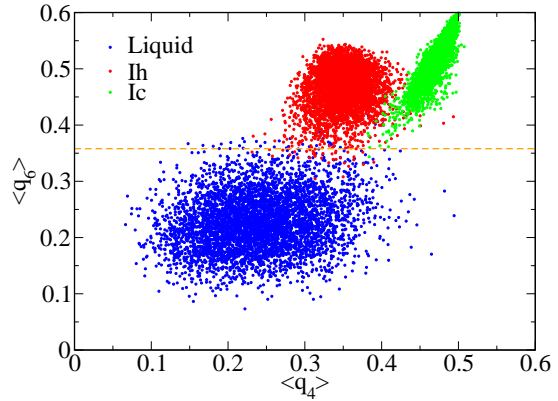}\\
\caption{Values of $\bar{q}_6$ and $\bar{q}_4$\cite{dellago} 
for 5000~molecules of the liquid phase (blue), of ice-Ih (red), and  of ice-Ic (green) 
at 237~K for the TIP4P/2005 model.} 
\label{q6q4}
\end{figure}

To determine the time evolution of the cluster size, we use the 
rotationally invariant order parameters proposed by Lechner and Dellago, $\bar{q_{i}}$.\cite{dellago}
In Fig.~\ref{q6q4} we show the $\bar{q}_{4},\bar{q}_{6}$ values for 5000~molecules of either liquid water, 
ice Ih or ice Ic at 1~bar and 237~K for TIP4P/2005. The cut-off distance  to identify neighbors  for the calculation
of $\bar{q_{i}}$ is $3.5$~\AA\ between the oxygen atoms. 
This approximately  corresponds to the position of the first minimum of the 
oxygen-oxygen pair correlation function in the liquid phase. 

From Fig.~\ref{q6q4} it is clear that $\bar{q}_{6}$ alone is enough to
discriminate between solid-like and fluid-like molecules, as already suggested
in Ref.~\onlinecite{doye_tip4p_2005}. As a threshold to separate the 
liquid from the solid clouds in Fig.~\ref{q6q4} we 
choose $\bar{q}_{6,t}=0.358$, represented as a horizontal
dashed line in the figure.  
This threshold separates the liquid from both ice Ih and Ic. Therefore, 
even though we prepare the clusters with ice-Ih structure, ice-Ic molecules would 
be detected as solid-like should they appear as the clusters grow.
Unlike Refs.~\onlinecite{ghiringhelli08} and \onlinecite{li13} we do not consider
as solid-like particles on the surface which are neighbor to solid-like particles. 
Once molecules are labelled either as
solid or liquid-like, the solid cluster is found by means of a clustering
algorithm that uses a cut-off of 3.5~\AA\ to find neighbors of the same
cluster. 

\subsection{Initial configuration}
\begin{figure}[h!]
\includegraphics[width=0.4\textwidth,clip=]{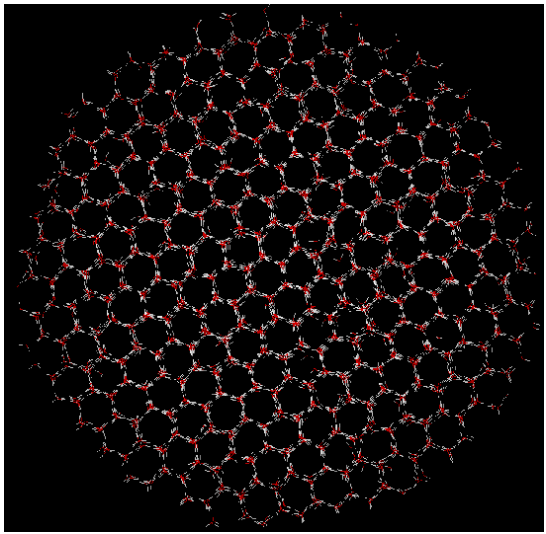}
\caption{Snapshot of a spherical ice-Ih cluster of 4648~molecules.} 
\label{cluster}
\end{figure}

We prepare the initial configuration by inserting a spherical ice-Ih cluster
(see Fig.~\ref{cluster} for a cluster of 4648~molecules) into a configuration
of supercooled water with $\sim 20$~times as many molecules as the cluster. 
To obtain the cluster, we  simply cut a spherical portion of a large
equilibrated  ice Ih crystal. 
Next, we insert the ice cluster in the supercooled liquid 
removing the liquid molecules that overlap with the cluster.
Finally, we equilibrate the system for about 0.2~ns at 200~K.
This time is long enough to equilibrate the cluster-liquid interface 
(see Supporting Information). 
We then perform simulations for three different system/cluster sizes labeled as H (Huge), L (Large) and B (Big) 
(see Table~\ref{systemsize}). As far as we are aware, the studied system size are 
beyond any previous numerical study of ice nucleation. Calculations were performed in the Spanish super-computer 
Tirant. 
For system H we use 150~nodes yielding 0.72~ns/day; 
for system L, 50~nodes at 1.5~ns/day and, 
for system B, 32~nodes at 4.7~ns/day. 

\begin{figure}[h!]
\includegraphics[width=0.2\textwidth,clip=]{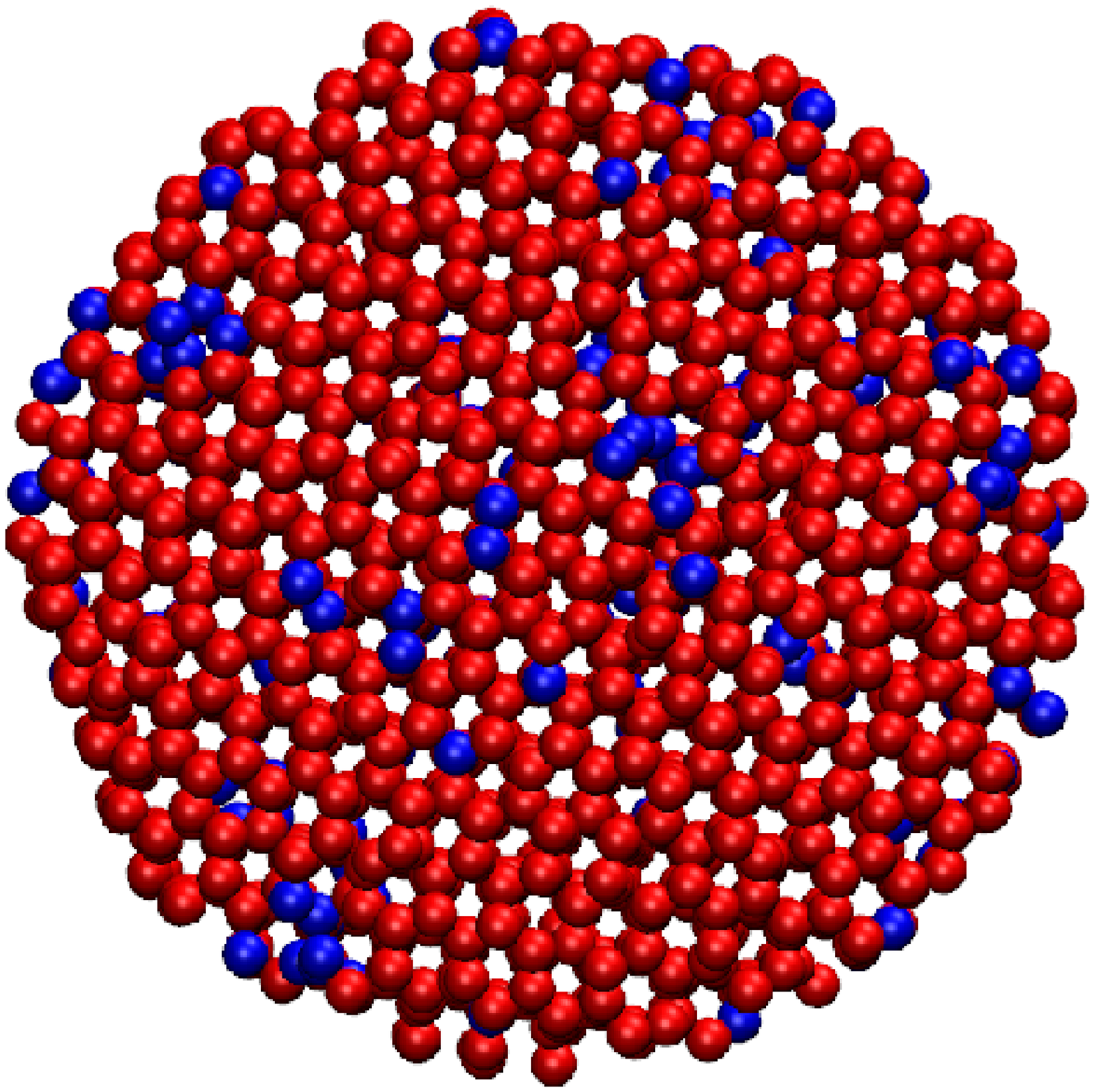}(a)
\includegraphics[width=0.2\textwidth,clip=]{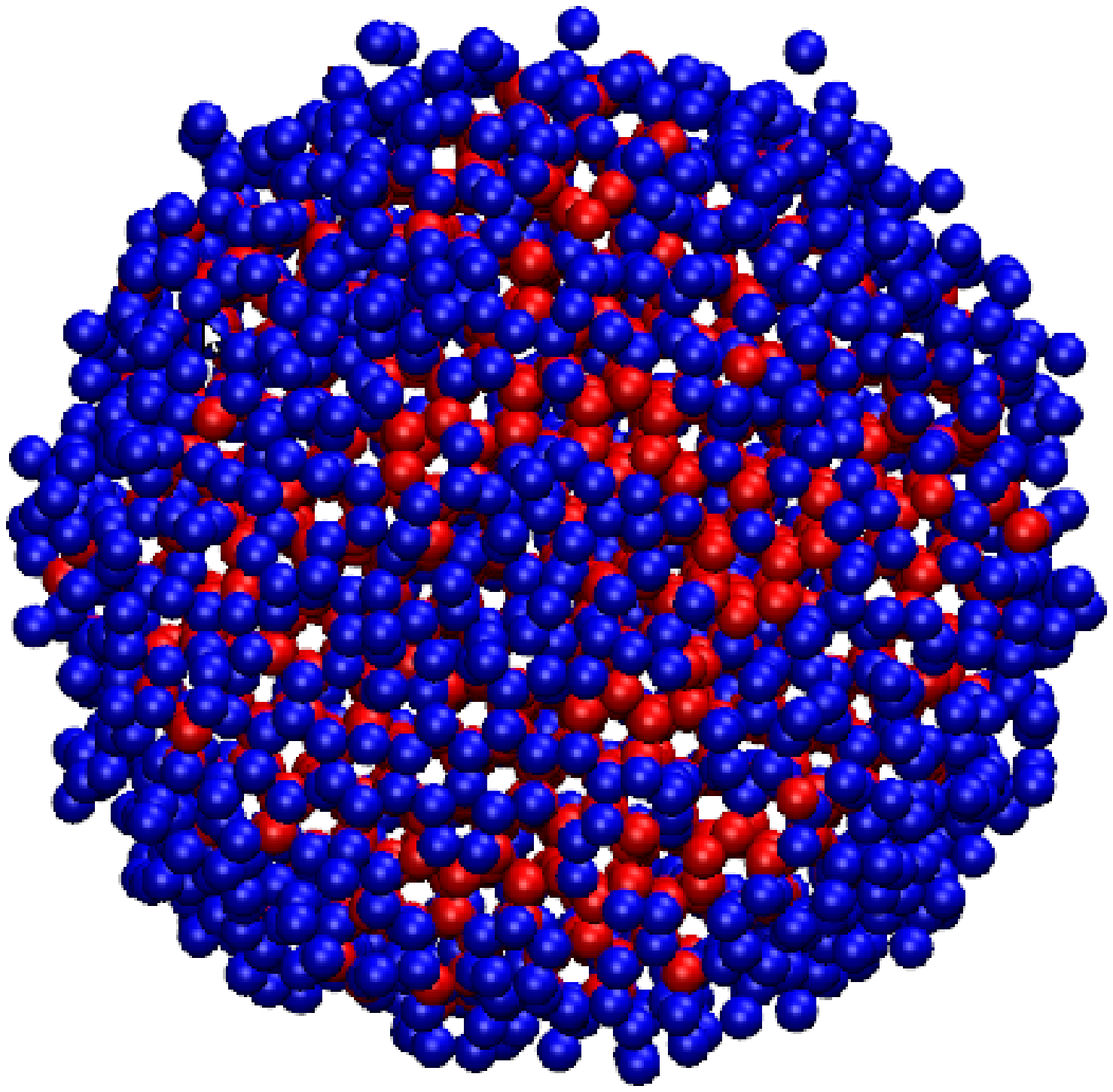}(b)\\
\includegraphics[width=0.2\textwidth,clip=]{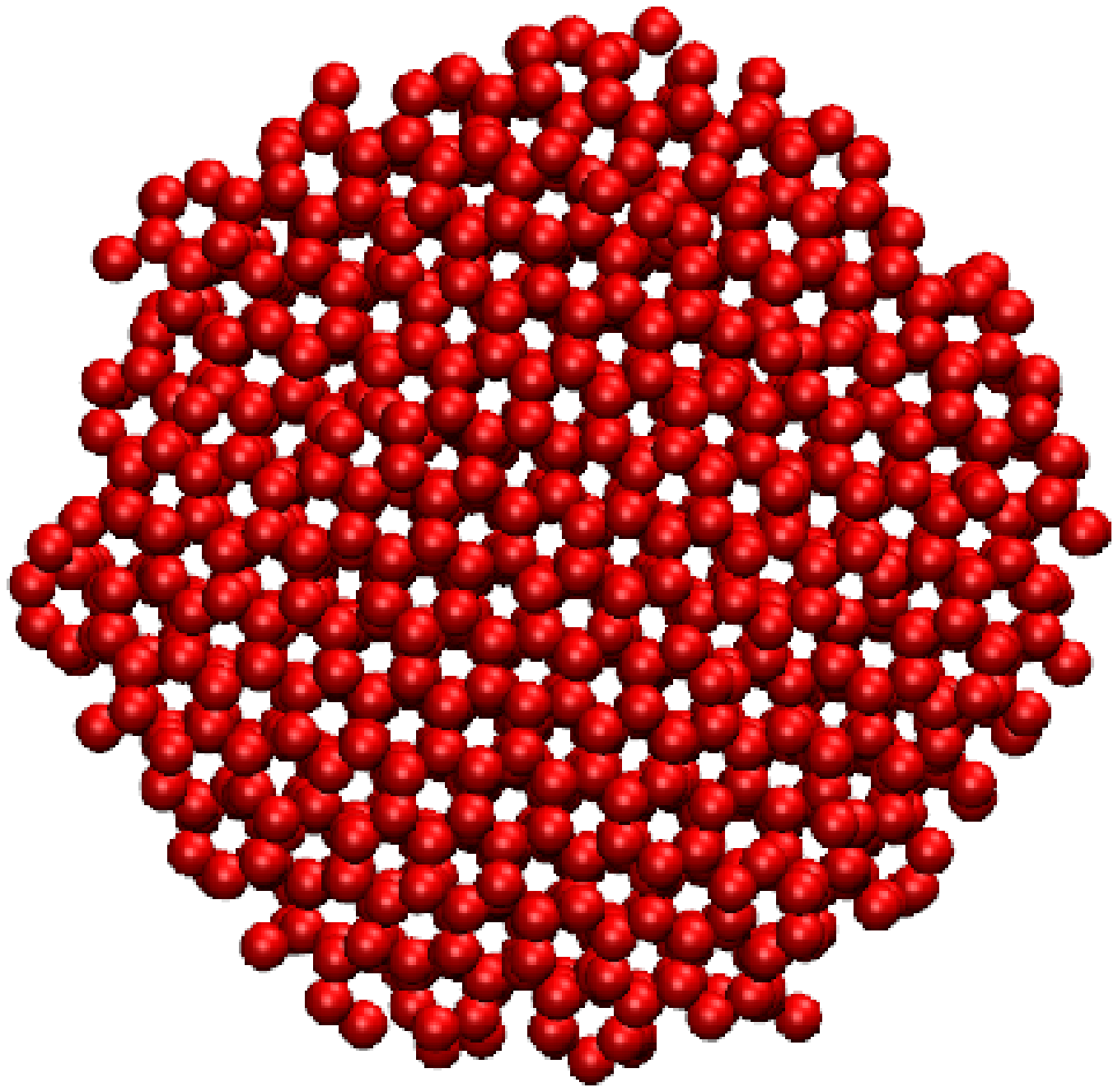}(c)
\caption{Snapshot of the 4648~molecules inserted as an ice cluster just after insertion (a),
and after 0.2~ns equilibration (b).  
Molecules are colored in red if detected as solid-like and 
in blue if detected as liquid-like. In (c) only solid-like molecules of snapshot (b) are shown.} 
\label{cluster-po}
\end{figure}

Our order parameter allows us to correctly identify as solid-like the great
majority of the molecules belonging to the cluster shown in Fig.~\ref{cluster}
(4498 out of 4648). 
Fig.~\ref{cluster-po}(a) shows that indeed most molecules of the inserted ice cluster are detected as solid-like (red) 
as opposed to liquid-like (blue). Notice that most blue particles in Fig.~\ref{cluster-po}(a) are located at the interface. 
This is not surprising giving that our order parameter was tuned to distinguish between liquid-like and solid-like particles in the bulk.  
Fig. ~\ref{cluster-po}(a) corresponds to the cluster {\em just} inserted in the liquid. 
After 0.2~ns of equilibration our order parameter detects that the number of molecules in the cluster drops down to 3170. 
To explain the origin of this drop we show in Fig.~\ref{cluster-po}(b) 
a snapshot of the 4648 inserted molecules after the 0.2~ns equilibration period. 
Clearly, the drop comes from the fact that the outermost layer of molecules of the inserted cluster becomes liquid-like during equilibration.
By removing the liquid-like molecules from Fig.~\ref{cluster-po}(b) one can easily identify again the hexagonal 
channels typical of ice (Fig.~\ref{cluster-po}(c)).    
Therefore, the drop from 4648 to 3170~molecules in the ice cluster is due to the equilibration of the ice-water interface. 
The size of the equilibrated clusters, $N_c$, is given in Table~\ref{systemsize}. 

\begin{table}[h!]
\caption{Total
number of molecules in the system, $N_t$ (ice cluster + surrounding liquid water molecules) 
and number of molecules of the inserted spherical ice cluster, $N_i$ for the 
three configurations prepared. $N_c$ is the number of molecules in the ice cluster after equilibration of the interface.  
The radius of the equilibrated clusters $r_c$ in \AA ~is also presented.}
\label{systemsize} 
\centerline{
\begin{tabular}{ccccccc}
\hline 
 System   &   $N_{t}$   &     $N_{i}$   & $N_{c}^{2005}$ &  $N_{c}^{Ice}$ & $r_{c}^{2005}$ & $r_{c}^{Ice}$  \\
\hline
B   & 22712  & 1089  & 		600  & 600 & 16.7 & 16.8 \\
L   & 76781  & 4648  & 		3170  & 3167 & 29.1 & 29.2  \\
H   & 182585 & 9998  & 		7931  & 7926 &  39.5 & 39.7 \\
\hline 
\end{tabular}}
\end{table}

Once the interface is equilibrated for $0.2$ ns, the number of molecules in the cluster grows
or shrinks (depending on the temperature) at a much slower rate (typically
requiring several nanoseconds as it is shown in Fig.~\ref{figlargest}). 
The initial time in our simulations corresponds to the configuration equilibrated after 0.2~ns. 
We run MD simulations of the system with the equilibrated interface at 
several temperatures below the bulk melting temperature of the model.
The objective is to find a temperature range within which the cluster
can be considered to be critical.
The temperature range is comprised between the lowest temperature at which the solid cluster
melts and the highest at which it grows.  
We monitor the number of molecules in the cluster 
and the global potential energy to find whether the cluster melts or grows.

\section{Results}

\subsection{Size of the critical clusters}

\begin{figure}[h!]
\includegraphics[width=0.45\textwidth,clip=]{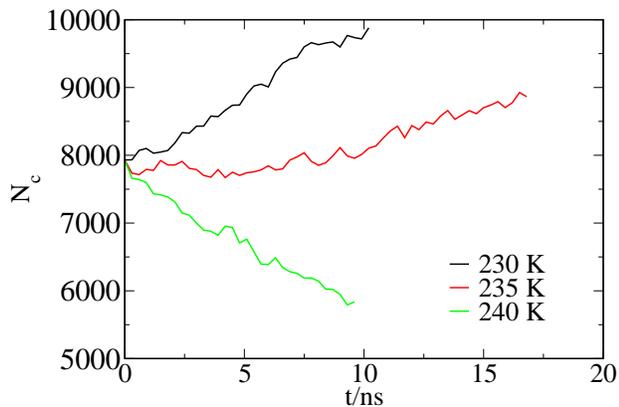}
\caption{Number of molecules in the ice cluster versus time for system H and the TIP4P/2005 potential. Results are shown for different temperatures as
indicated in the legend.} \label{figlargest}
\end{figure}

In Fig.~\ref{figlargest} we represent the number of molecules in the ice cluster
versus time for system H, TIP4P/2005. 
Depending on the temperature the cluster either grows (230~K and 235~K) or
shrinks (240~K).  The highest temperature at which the cluster grows is 235~K
and the lowest temperature at which it melts is 240~K. Hence, a cluster of
$\sim$7900~molecules (as detected by our order parameter) is critical at
$237.5 \pm 2.5$~K. 
An analogous result can be obtained by monitoring the potential energy of the
system as a function of time.
(see Supporting Information).
A decrease in the energy corresponds to the cluster's growth 
whereas an increase in the energy corresponds to its melting. 
By doing this analysis for both models (TIP4P/2005 and TIP4P/Ice ) and for the
three cluster sizes (H, L, and B ), 
we obtain the results summarized in Table~\ref{tabcluster}. 

\begin{table}[h!]
\caption{
We report the
temperature ($T$ in K) for which the prepared ice clusters are found to be  critical,  
the supercooling ($\Delta T$ in K) for the corresponding water model, 
the chemical potential difference between the fluid and the solid ($\Delta \mu$ in kcal/mol), 
the liquid-solid surface free energy ($\gamma$ in mN/m) estimated from Eq.~\ref{ncrit},
and the nucleation free energy barrier height ($\Delta G_{c}$ in $k_B T$) estimated from Eq.~\ref{eq_G_cnt}.  
\label{tabcluster}} 
\centerline{
\begin{tabular}{ccccccccc}
\hline 
  Model & System         & $N_c$ & $T$  & $\Delta T$ & $\Delta \mu$ & $\gamma$ & $\Delta G_{c}$\\
\hline
TIP4P/2005 & B &  600  & 222.5  & 29.5   &  0.114      &   20.4    &   77 \\
TIP4P/2005 & L &  3170 & 232.5  & 19.5   &  0.080      &   24.9    &  275 \\
TIP4P/2005 & H &  7931 & 237.5  & 14.5   &  0.061      &   25.9    &  515 \\
\hline
TIP4P/Ice & B  & 600  & 237.5   & 34.5   &  0.133      &   23.6    &   85 \\
TIP4P/Ice & L  & 3167 & 252.5   & 19.5   &  0.083      &   25.4    &  261 \\
TIP4P/Ice & H  & 7926 & 257.5   & 14.5   &  0.063      &   26.3    &  487 \\
\hline 
\end{tabular}}
\end{table}

\begin{figure}[h!]
\includegraphics[width=0.45\textwidth,clip=,angle=0]{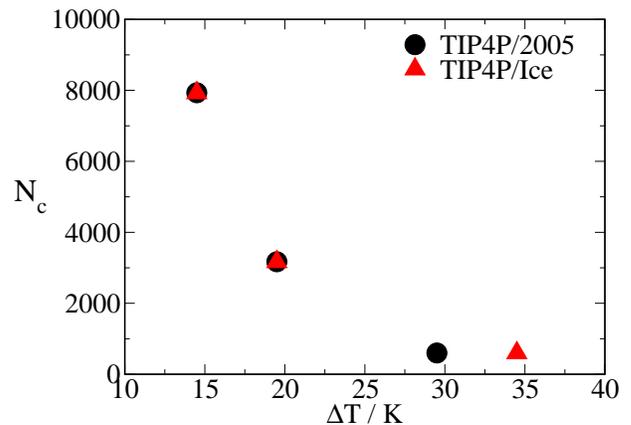}
\caption{Critical cluster size versus $\Delta T$ for the studied water models.
Notice that the points corresponding to both models at low supercooling are essentially on top 
of each other.}
\label{masterfigure}
\end{figure}

For the temperatures explored in this work (from about 15~K to 35~K below the
melting temperature of both TIP4P/2005 and TIP4P/Ice) the size of the ice
critical cluster ranges from nearly 8000 (radius of 4 nm) to about 600 molecules (radius of 1.7 nm). 
This compares reasonably well with a 
critical cluster radius of $\sim$ 1.3 nm obtained by 
applying CNT to 
experimental measurements at a supercooling of about 40 K.\cite{micelas,kiselev01} 
\edu{Our results are also consistent with CNT based estimates of the critical size at lower
supersaturations.\cite{bogdan,kashchievbook} For instance, in Fig. 
15.7 of Ref. \cite{kashchievbook}, a critical
cluster size ranging from 1000 to 300 molecules is predicted for 25 K $<\Delta T <$ 30 K.} 
An interesting remark is
that the temperatures of the TIP4P/Ice are basically shifted 20~K above the
the corresponding ones for TIP4P/2005 with the same nucleus size. 
This is precisely the difference between the melting temperatures of both
models and, thus, the supercoolings are very similar for a given ice cluster size in both models.
This is more clearly shown in Fig.~\ref{masterfigure} where the size of
the critical cluster is plotted as a function of the the difference between the
melting temperature of the model and the temperature of interest $\Delta T=T_m-T$.
We observe that, within our error bar, the critical cluster size of both models 
scales in the same way with respect to their melting temperatures. This is not so surprising since 
TIP4P/2005 and TIP4P/Ice present a similar charge  distribution and mainly  differ in 
the choice of the potential parameters.

In previous works\cite{vega07,vega09,vega11} we observed that, for a number of properties, 
the values of TIP4P/2005 lie in the middle of the values obtained for TIP4P and TIP4P/Ice.
Therefore, it is expected that TIP4P gives similar results to TIP4P/2005 and TIP4P/Ice regarding
the dependence of $N_c$ with $\Delta T$. 
Matsumoto et al.\cite{matsumoto02} studied ice nucleation at 230~K and a density of 0.96~g/cm$^3$ using the
TIP4P model. This thermodynamic state point corresponds to a pressure of about -1000~bar and $\Delta T$ 5 K.\cite{sanz_prl} 
By extrapolating the data of Fig.~\ref{masterfigure} to $\Delta T=5 K$
one gets a critical cluster of the order of hundreds of thousand molecules. 
Therefore, 
it is likely that the results obtained by Matsumoto et al.\cite{matsumoto02}, although pioneering 
and useful to learn about the 
ice nucleation pathway, may suffer from system size effects and may not be valid
to estimate either the size of the critical cluster or the nucleation rate.

In an important paper, Koop et al.\cite{koop_nature} showed that the
homogeneous nucleation rate (and therefore the temperature of homogeneous
nucleation) of pure water and of water solutions can be described quite well by
a function that depends only on the water activity.
This conclusion has been confirmed in more recent experiments.\cite{knopf11}
Although the nucleation rate for an aqueous solution is the same as for pure
water, the freezing points are different. One is then tempted to suggest that
the size of the critical cluster at the homogeneous nucleation temperature
could be the same for pure water and for aqueous solutions. Moreover, the fact
that thermodynamics is sufficient to predict the rate seems to indicate that
the water mobility is also determined by the free energy of water. A
microscopic study of the relationship between crystallization rates, structure
and thermodynamics of water which may explain the empirical findings of Koop
and coworkers has recently been presented in Ref.~\onlinecite{nature_valeria}.

\subsection{Interfacial free energy and free energy barrier}

Once the size of the critical cluster is known, 
one can use Eq.~\ref{ncrit} to estimate the solid-liquid interfacial free energy. 
Since ice density changes little with temperature\cite{mendu_eos}, the density at coexistence is used in our calculations
($\rho_{m,TIP4P/Ice}$=0.906 g/cm$^3$ and $\rho_{m,TIP4P/2005}$=0.921 g/cm$^3$). 
For most substances it is possible to approximate $\Delta \mu$ by $\Delta h_m
(T_m-T)/T_m$, where $\Delta h_m$ is the melting enthalpy and $T_m$ is the
melting temperature. For water, however, this may not be a good approximation because $\Delta
h$ significantly changes with temperature as a manifestation of the anomalous
sharp increase of the  heat capacity of water as temperature
decreases.\cite{jeffery1997,Kumar05062007} Hence, one needs to do a proper
evaluation of the chemical potential difference between both phases to get the
surface free energy from Eq.~\ref{ncrit}. 
We have calculated $\Delta \mu$ at every temperature by means of standard
thermodynamic integration \cite{frenkel96} from the coexistence 
temperature, at which $\Delta \mu = 0$. 
In Table~\ref{tabcluster} we report the values we obtain for $\Delta \mu$
and $\gamma$. 

First of all we note that $\gamma$ decreases with temperature for both models.
This is in qualitative agreement with experimental estimates of the behavior
of $\gamma$ with $T$.\cite{pruppacher1995,zobrist07,alpert11} A more quantitative comparison is
not possible in view of the large discrepancies between different estimates
(see Fig.~10 in Ref.~\onlinecite{pruppacher1995}).  
Motivated by the fact that the interfacial free energy can only be measured at
coexistence, we extrapolate our results to the melting temperature. 
To do that, we take the two largest clusters and evaluate the slope of
$\gamma(T)$.
We get a value for the slope of $\sim$0.18~mN/(m~K) for both models, in
very good agreement with a recent calculation for the TIP4P/2005
model.\cite{doyejcp2013} 
With a linear extrapolation we get a value for $\gamma$ at $T_m$ of
$\sim$28.7~mN/m for both models, which can be compared to experimental
measurements.
In contrast with the vapor-liquid surface tension, the value of
$\gamma$ for the solid-fluid interface is not well established.
Experimental values range from 25 to 35~mN/m.\cite{pusztai2002} 
Our calculated data for $\gamma$ at coexistence lies in the middle of that
range, so our models predict a surface free energy which is consistent with
current experimental data.
We now compare our estimated $\gamma$ to direct calculations from simulations
using a planar interface.  The value of $\gamma$ depends on the plane in
contact with the liquid. Since the cluster used here is spherical 
we shall compare with the average of the values obtained for the basal and prismatic planes. 
Davidchak et al. computed $\gamma$ for a planar fluid-solid interface using two models similar to those
used in this work: TIP4P and TIP4P-Ew.
For TIP4P, in an initial publication the authors reported a value of  $\gamma
= 23.9$~mN/m\cite{gammadavid_old} that was later on modified (after improving
their methodology) to $\gamma = 26.5$~mN/m.\cite{gammadavid} For the TIP4P-Ew\cite{horn04}
Davidchak et al. reported (using the improved methodology) a value of
$27.6$~mN/m.\cite{gammadavid} TIP4P-Ew is known to predict water properties in
relatively close agreement to those of TIP4P/2005. Therefore, our results are also
consistent with the calculations reported in the literature for similar models.
To conclude, our values of $\gamma$ seem to be
reasonable estimates of the interfacial free-energy of the planar ice-water
interface.

To estimate the height of the nucleation free-energy barrier 
we make use of Eq.~\ref{eq_G_cnt}.
Our results are summarized in Table~\ref{tabcluster}. 
In view of the height of the nucleation barrier for the clusters of systems
L and H, around 250 and 500 $k_B T$ respectively, it seems virtually impossible
to observe homogeneous nucleation of ice for supercoolings lower than 
20~K.  
The height of the nucleation barrier provides an estimate of the 
concentration of critical clusters in the metastable fluid as 
$\rho_f \exp(-\Delta G_{c}/(k_B T))$, where $\rho_f$ is the number density 
of the fluid. 
For $\Delta G_{c} = 250$~$k_B T$, one critical cluster would appear on average
in a volume $\sim$10$^{60}$ times larger than the volume of the whole
hydrosphere.  From the values of $\Delta G_{c}$ of  Table~\ref{tabcluster}  we may infer why 
spontaneous ice nucleation has never been observed in previous studies of
supercooled water with the TIP4P/2005 model\cite{abascal10,abascal11,pettersson11}. Our 
results show that the free energy
barrier for nucleation even for temperatures as low as 35~K below melting is
still of about 80 $k_B T$. This is much larger than  the typical barrier found in studies where spontaneous 
crystallization occurs in brute force simulations \cite{hs_filion,lundrigan:104503} (about $18 k_B T$). 
It is worth mentioning that neither Shevchuk and  Rao\cite{rao} nor Overduin and Patey\cite{patey_sc} 
find any evidence of ice nucleation in TIP4P models after runs of several microseconds which is consistent with the
results of this work. 
Our results may be of great interest to studies in which the competition between the 
crystallization time and the equilibration time of water is crucial\cite{limmer13,poole13,liu12}.

\subsection{Nucleation rate}

Although the free energy barriers alone provide a strong indication that ice
can not appear on our planet  via homogeneous nucleation at moderate
supercoolings ($\Delta T < 20$~K), it is worth calculating the nucleation rate,
$J$, to confirm such statement. 
The nucleation rate takes into account not only the concentration of the 
clusters but also the speed at which these are formed. 
Moreover, the supercoolings for the smallest clusters we investigate are
comparable to those where most experimental measurements of $J$ have been made
($\Delta T \sim$~35~K).\cite{taborek,pruppacher1995,knopf11,riechers13,pruppacher_book}

\begin{figure}[h!]
\includegraphics[width=0.45\textwidth,clip=,angle=0]{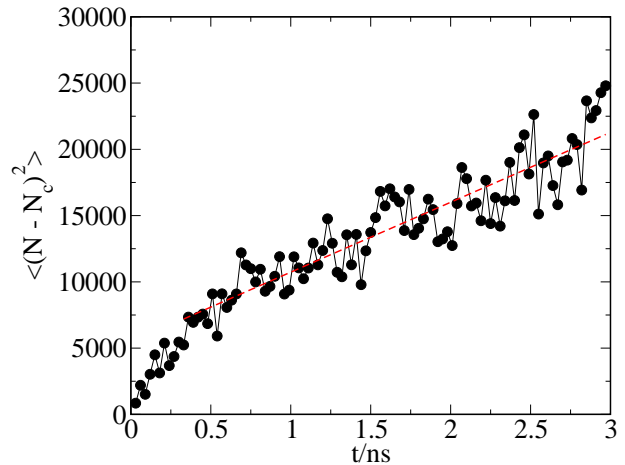}
\caption{$<(N(t)-N_c)^2>$ versus time for configuration L, TIP4P/2005. The  attachment rate $f^{+}$ is obtained as half the
value of the slope. The curve above is obtained as an average over 30 trajectories. 
In approximately half of these trajectories the critical 
cluster ended up growing, whereas it eventually melted in the other half. 
}
\label{attach}
\end{figure}

To calculate the nucleation rate we use Eq.~\ref{eqrate}. 
First, we compute $f^{+}$ from Eq.~\ref{eqattach} by running 30 simulations of
the cluster at the temperature at which it was determined to be critical. We
monitor $(N(t)-N_c)^2$ and average it over all the runs.  In Fig.~\ref{attach}
we plot $<(N(t)-N_c)^2>$ versus time for the system L, TIP4P/2005.
From the slope at long times we can infer $f^{+}$.\cite{auerjcp} We get 
$f^{+} = 70\cdot10^9$~$s^{-1}$. The Zeldovich factor for this particular case is $1.77\cdot10^{-3}$, 
and the density of the liquid is 0.977~g/cm$^3$. With this, we have all the ingredients needed
to calculate the nucleation rate via Eq.~\ref{eqrate}. The final result for this 
case is $\log_{10} (J /$(m$^{-3}$s$^{-1}$))=-83. 

\begin{figure}[h!]
\includegraphics[width=0.45\textwidth,clip=,angle=0]{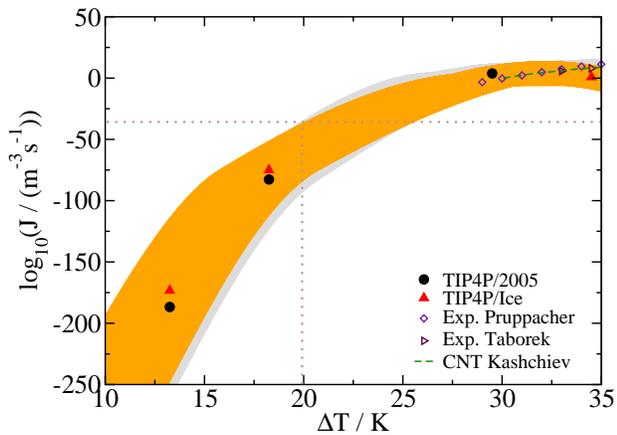}
\caption{Nucleation rate as a function of the supercooling.
\edu{Symbols correspond to our simulation results and to 
experimental measurements as indicated in the legend. The green dashed line corresponds
to CNT estimates of $J$.\cite{kashchievbook}} The grey and orange shadows 
represent the estimated error bars for TIP4P/2005 and TIP4P/Ice respectively interpolated
by splines. The horizontal \edu{dotted} line indicates the rate given by the growth of one cluster in
the age of the universe in all the water of the Earth's hydrosphere. The vertical 
\edu{dotted} line indicates the supercooling below which homogeneous nucleation is impossible.}
\label{rate}
\end{figure}

The same procedure is used to calculate the nucleation rate for the rest of the
systems described in Table~\ref{systemsize}. The results for the nucleation
rate as a function of the supercooling are presented in Fig.~\ref{rate}
and compared to the experimental measurements of Pruppacher\cite{pruppacher1995}
and Taborek\cite{taborek}.
The horizontal dashed line corresponds to the nucleation rate required for the
appearance of one critical cluster in the volume of Earth's hydrosphere in
the age of the universe, which we call ``impossible nucleation rate''.
The vertical line shows at which temperature the impossible nucleation rate
line intercepts the upper limit of our error bars (grey and orange shadows for
TIP4P/2005 and TIP4P/Ice respectively). 
In view of this figure we can confidently claim what the free energy barriers
previously hinted: it is impossible that ice nucleates homogeneously in our
planet for $\Delta T < 20$~K. 
In other words, heterogeneous nucleation must take place in order for water to
freeze for supercoolings lower than 20~K. This is consistent with the fact
that, when heterogeneous nucleation is suppressed, moderately supercooled water
can remain  metastable long enough for its thermodynamic properties to be
measured.\cite{speedy76,hare_sorensen_eos_supercooled_water,mishima10,debenedetti03,holten12} 
From our results it is also clear that ice formation should not be expected in
brute force molecular dynamics simulations at moderate supercoolings (provided
that the system is large enough not to be affected by finite size
effects).\cite{matsumoto02} To  observe  ice formation in brute force
simulations the nucleation rate should be higher than $\log_{10} (J
/$(m$^{-3}$s$^{-1}$)) = 32 (this number is obtained assuming the formation of
ice after running about 100~ns in a system of about 50nm$^3$, which are typical
values in computer simulations of supercooled water).  Notice also that the
maximum in the isothermal compressibility at room
pressure\cite{abascal10,abascal11} found at about $\Delta T=20$~K for the
TIP4P/2005 model can not be the ascribed to the transient formation of ice
as the nucleation rate of ice at this temperature is negligible. 

Another interesting aspect of Fig.~\ref{rate} is the comparison with
experiment.  Both models give nucleation rates that reproduce the experimental
measurements within the uncertainty of our method.  This excellent result
brings confidence in  the ability of the selected models to predict relevant
quantities for the nucleation of ice such as the nucleation rate, the critical
cluster size, and the surface free energy. 

\edu{We also include in Fig.~\ref{rate}
a green dashed line that corresponds to
the CNT based estimates of $J$ shown in Fig. 13.6 of Ref. \cite{kashchievbook} The agreement 
between CNT, simulations and experiments is quite satisfactory.   
To the best of our knowledge, there are no CNT estimates of $J$ available for
supersaturations lower than 30 K to compare our results with.\cite{pruppacher1995,kashchievbook,murray2010}} 

By using Forward Flux Sampling\cite{PRL_2005_94_018104}, Li {\em et al.}
determined $J$ for the mW model of water for temperatures between 35 and 55~K
below the model's melting temperature.\cite{galli_mw} Since we are
interested in ice nucleation at moderate supersaturation, our study deals with
lower supercoolings 
($14.5$~K$ < \Delta T < 34.5$~K). 
Nonetheless, our highest supercooling (34.5~K) is very close to the lowest one of  
Li {\em et al.} (35~K) so we can compare both results. 
The values of Li {\em et al.} for $J$ are 5-8 orders of
magnitude below the experimental ones when compared at the same absolute
temperature (the deviation increases when the comparison is made at the same degree of
supercooling). The nucleation rates calculated in this work for TIP4P/2005 and
TIP4P/ice are similar (although slightly larger) to those for the mW model. Initially this may
appear surprising as the mW model is a coarse grained model of water with no
hydrogens, which makes its dynamics faster than that of both real water and TIP4P-like models.\cite{mw}
However the free energy barrier of mW may be larger, compensating this kinetic effect. 
In fact the interfacial free energy of mW has been found\cite{galli_mw} to be $\gamma=31$~mN/m 
(larger than the values found in this work for TIP4P/2005 and TIP4P/Ice).
This high value of $\gamma$ may be partially compensated by a significant overestimate
of the ice density of this model (0.978 g/cm$^3$ to be compared to the experimental result 0.91 g/cm$^3$).
The net balance is that the values of $J$ of the mW model are similar,
although somewhat lower, than those for TIP4P/2005 and TIP4P/ice.

As for the size of the critical cluster, we find that it is of about 600
water molecules for TIP4P/Ice at 237.5~K ($\Delta T = 34.5$~K). 
Li {\em et al.} have reported a critical cluster size of about 850
molecules for the mW model at 240~K ($\Delta T = 35$~K).
Both results are compatible since Li {\em et al.} include in the ice cluster molecules which 
are neighbor to the solid cluster, and we do not. 
In summary, our results for TIP4P/2005 and TIP4P/Ice 
are consistent with Li's {\em et al.} for mW. 


\section{Discussion}
\subsection{Validity and possible sources of error}
\label{validity}

The methodology we have used is subject to two main error sources: the determination
of the cluster size and the location of the temperature at which the clusters
are found to be critical. Moreover, our approach relies on the validity of 
CNT. In the following paragraphs we discuss the extent to which our 
results may be affected by these issues.  

In nucleation studies, the size of the largest solid cluster is usually
considered a good reaction coordinate.
To identify the cluster, we first need to distinguish between liquid-like and solid-like molecules.
The chosen criterion should be able to identify the majority of molecules of the bulk solid as solid-like, 
and the majority of molecules of the  bulk fluid  as liquid-like. 
One could in principle find several criteria  that successfully perform this task. 
However,  when interfaces are present in the system 
(as in the case of a solid-liquid\cite{hs_filion} 
or a solid-vapor\cite{maria06b,conde08} interface) depending on the chosen criterion 
one might assign differently the interfacial molecules (see for instance 
References~\onlinecite{doye12} and \onlinecite{galli_mw} 
for an illustration of this problem for the mW water model).  

How does the choice of a criterion to distinguish liquid from solid-like molecules affect our results? 
Whether the cluster grows or shrinks for a given temperature does
not depend on the particular choice of the order parameter (see Supporting Information).
The same trend can be obtained by monitoring 
global  thermodynamic properties of the system, such as the total potential energy (see Supporting Ingormation). 
Therefore,  the fact that the cluster shown in Fig.~\ref{cluster-po} 
is critical at 232.5~K is independent on the particular choice of the  criterion to distinguish liquid from solid-like molecules.  

A different problem arises if one asks the question: how many ice molecules  are 
present in Fig.3b? Different criteria provide different answers  even though the configuration 
presented in Fig.3b is unique. 
Since the origin of this arbitrarity is due to the interfacial region, it is expected that
the arbitrarity will become smaller as the ice cluster becomes larger. 
However, for the system sizes considered in this work the interface region still matters. 
To take this effect into account we have estimated the error bars in Fig.~\ref{rate}
considering an arbitrarity of 60\% in the labeling of {\em interfacial} molecules. 
This would affect the value of $\gamma$ by  7\%, and the free-energy barriers height 
by up to 20\%. 
Although this estimated error seems large, it is worth pointing out that differences between the free-energy barrier estimated by
different groups may be, in the case of water, much larger than that.\cite{trout_prl_2003,trout03,quigley08,anwar_tip4p_nucleation} 
In summary we  conclude that the liquid/solid criterion chosen in this work provides reasonable estimates of 
$\gamma$, and when used within the CNT framework allows to interpret our simulations results 
in a rather straightforward way. 

Another important error source in the calculation of $J$ is the location of the temperature 
at which a cluster is critical. As we show in Fig.~\ref{figlargest}, by performing runs at different temperatures we identify, within a certain
range, the temperature that makes critical a given ice cluster. 
We assign the temperature in the middle of the range to the corresponding cluster, but the temperature
that really makes the cluster critical 
could in principle be any other within the range. 
This uncertainty 
has a strong contribution to the error bars in Fig~\ref{rate}, particularly at low supercoolings,
where the variation of $J$ with $T$ is very steep.  
This error could, in principle, be easier to reduce than that coming from the arbitrarity in the determination
of the number of particles in the cluster. One simply has to do more runs to narrow 
the temperature range. However, these simulations are very expensive given the large system 
sizes we are dealing with. 
It is interesting to point out that temperature control is also seen as a major
error source in experiments.\cite{riechers13} 

Our results for $\gamma$, $\Delta G_c$, and $J$ rely on the validity of CNT. 
Classical Nucleation Theory is expected to break down for small clusters, when 
the view of nucleation as a competition between bulk and surface free energies starts to be
questionable (in clusters of a few hundred particles most molecules are placed at
the surface). However, for the large cluster sizes investigated in this work 
it seems reasonable to assume that CNT works well. The satisfactory comparison 
of our estimate of $\gamma$ with that obtained in simulations of a flat interface\cite{gammadavid}
is certainly encouraging in this respect. 
Moreover, we have applied the methodology described in this paper 
to calculate the nucleation rate of the mW water model
and we get, within error, the same nucleation rate as in Ref.\cite{galli_mw}  This is a very
stringent test to our approach, given that in Ref\cite{galli_mw} 
a method that relies neither on CNT nor on the definition of the
cluster size was used (Forward Flux Sampling). This comparison is made for a supercooling of 35 K, the deepest
investigated in this work. 
For lower supercoolings, where the critical cluster is larger,
the methodology is expected to be even more robust.
The advantage of the approach used here is that it allows to estimate (at a 
reasonable computational cost) critical cluster sizes and 
nucleation rates at low and moderate supercooling.

\subsection{Novelty}

In this paper
we provide values for the homogeneous nucleation rate of ice at moderate 
supercoolings ($\Delta T < 33$~K). For the first time, this is done without 
extrapolating from measurements at high supercoolings.
The experimental determination of $J$ is limited to a narrow temperature window 
at high supercoolings
(between 233~K and 239~K). In that window, $J$ can be directly measured 
without introducing any type of approximation. 
It only requires the knowledge of the droplet volume, the cooling
rate and the fraction of freezing events.  
Differences in the value of $J$ 
between different experimental groups are
relatively small (between one and two orders of magnitude). 
Therefore, the experimental
value of $J$ is well established for the narrow range of temperatures in which the current experimental
techniques can probe the nucleation rate.\cite{taborek,pruppacher1995,knopf11,riechers13,pruppacher_book}
To obtain values of $J$ outside that temperature window 
one can either extrapolate the data or make an estimate via CNT. 
An extrapolation from such a narrow temperature window would not be very reliable because  $J$ changes sharply with T. 
In turn, an estimate of $J$ based on CNT relies in the knowledge of the interfacial free energy. 
Unfortunately, our current knowledge of $\gamma$ for the water-ice
interface is far from satisfactory in at least three respects. Firstly, the
calculated values of different groups using CNT differ
significantly (see for instance Tables I and II in Ref~\onlinecite{murray2010}). 
Secondly, the values obtained for $\gamma$ from CNT seem to
be different from those determined for a planar ice-water interface at the
melting point (see for instance Fig.~8 in Ref~\onlinecite{bartell94}).
Finally, there is even no consensus about the
value of $\gamma$ for a planar interface at the melting point of water, a
magnitude that in principle could be obtained from direct experiments without
invoking CNT (values between 25 and 35~mN/m have been reported). A look to
Fig.~10 of the classic paper of Pruppacher\cite{pruppacher1995}
is particularly useful. It shows the enormous
uncertainty that exists at any temperature about the value of $\gamma$ for the
ice-water interface. Since $\gamma$ enters in the estimation of $J$ as
a power of three in an exponential term, the enormous scatter implies that,
at this moment, there is no reliable estimate of the value of $J$ for moderately supercooled water arising from
CNT.
In other words, you can get many different estimates of $J$ from the different
estimates of $\gamma$ shown in the paper by Pruppacher. \edu{In addition, to the best of our knowledge, 
no one has estimated $J$ using CNT for supersaturations lower than 30 K. \cite{pruppacher1995,kashchievbook,murray2010}}

\edu{Regarding the critical nucleus size}, it is not possible at the moment to 
measure it experimentally
by direct observation. 
Therefore, the prediction of the critical cluster at moderate and experimentally accessible supercoolings
is a novel result. 
Since the TIP4P/2005 has been quite successful in describing a number of
properties of water (notably including the surface tension for the vapor-liquid
equilibrium) we believe that the values reported here for $\gamma$ and $J$ from our
analysis of the critical cluster are a reasonable estimate for the
corresponding values for real water.

\subsection{Summary and outlook}

We have studied homogeneous ice nucleation by means of computer simulations 
using the TIP4P/2005 and TIP4P/Ice  water models. 
This is the first calculation of the size of the critical cluster and 
the nucleation rate at moderate supercoolings (14.5-35~K). 
Both models give similar results when compared at the same supercooling. 

To determine the size of the critical cluster, we use a numerical approach in
the spirit of direct coexistence methods.
We prepare an initial configuration by inserting a large ice cluster (about
10000, 4600 and 1000~molecules) in an equilibrated sample of liquid water.
Then, we let the interface equilibrate for 0.2 ns at 200~K.  Finally, we
perform molecular dynamic runs at several temperatures to detect either the
melting or the growth of the inserted cluster by monitoring its size.  We find
that the size of the critical cluster varies from $\sim$8000~molecules
(radius$ = 4$~nm) at 15~K below melting to $\sim$600 molecules (radius$ =
1.7$~nm) at 35~K below melting.  

We use CNT to estimate
the interfacial free energy and the nucleation free energy barrier.
Our predictions show that 
the interfacial free energy decreases as the supercooling increases, in agreement
with experimental predictions. An extrapolation of the interfacial free energy to the melting
temperature gives a value of 29(3)~mN/m, which is in reasonable agreement 
with experimental measurements
and with estimates obtained from computer simulations for TIP4P-like models.
We get free energy barriers higher than 250~kT
for supercoolings lower than 20~K. This strongly suggests that homogeneous ice nucleation
for supercoolings lower than 20~K is virtually impossible. 
We confirm this by calculating the nucleation rate. To do that we compute, by means of molecular 
dynamics
simulations,  
the rate at which particles attach to the critical clusters. 
These calculations show that, indeed, for supercoolings lower than 20~K it is impossible
that ice nucleates homogeneously. According to this prediction, ice nucleation must necessarily 
be heterogeneous for supercoolings lower than 20~K. The nucleation rate we obtain at higher 
supercoolings (30-35~K) agrees, within the statistical uncertainty of our methodology, 
with experimental measurements. 

It would be interesting to extend this work in several directions. Modifying the shape of the inserted 
cluster (inserting for instance a small crystal with planar faces) or even inserting a block of 
cubic ice Ic to analyse whether this cluster may be more stable as suggested by some studies\cite{huang_bartell,murray2010} 
are interesting issues that deserve further studies. 
Secondly, it would be of interest to consider other water models, to analyse the possible similarities/differences
with respect to nucleation of different potential models varying significantly either in the charge distribution 
as TIP5P\cite{mahoney00} or in the way the tetrahedral order is induced as in the mW model.\cite{mw}
Analyzing the behaviour at higher degrees of supercooling than those presented here is another interesting problem as 
well as the determination of the growth rate of ice.\cite{kusalik_growth_ice}
We foresee that all these issues  will be the centre of significant activity in the near future. \\ 

\section{Supporting Information to ``Homogeneous ice nucleation at moderate supercooling from molecular simulation"}
\subsection{Equilibration of the initial configuration}

\begin{figure}[h!]
\centerline{\includegraphics[clip,width=0.5\columnwidth]{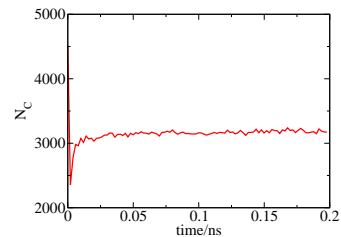}}
\caption{Equilibration for the L system of the TIP4P/2005 model. After an initial drop of the cluster size due to equilibration of the interphase, the cluster size changes very slowly with time.}
\label{fig:purity}
\end{figure}

The preparation protocol used  for all cluster sizes is the following.
After having inserted the ice cluster in the supercooled liquid, we remove the liquid molecules 
overlapping with the solid ones. Next, we equilibrate the system for about 0.2~ns at 200 K.
To make sure that the chosen 0.2~ns is a proper equilibration time, long enough to allow 
for annealing mismatches at the interface, we run a simulation 
starting from the initial configuration at time zero. 
In Fig.~\ref{fig:purity}, we represent the cluster size for the Large system of
the TIP4P/2005 model as a function of time.  
The figure shows that, even though the dynamics is very slow at such low
temperatures,  the chosen equilibration time is long enough. This result is
independent on the chosen cluster size or water model potential.

\subsection{Choice of the order parameter to distinguish between liquid/solid particles}

\begin{figure}[h!]
\centering
\includegraphics[width=0.5\columnwidth]{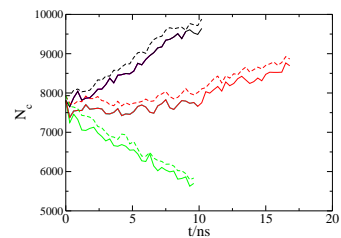}
\caption{Number of particles in the cluster versus time for system size H, TIP4P/2005 model. Black curves correspond to 230 K, red ones to 235 K and green to 240 K. 
Solid lines correspond to the analysis made with the order parameter described in the main text. Dashed lines correspond to the use of an alternative order parameter.  
With such order parameter particles are considered as neighbors if their oxygen atoms are closer than 3 ~\AA\ and are labelled as solid-like whenever 
their $\bar{q}_{3}$ is larger than 0.28.}
\label{q3op}
\centering
\end{figure}

The use of an alternative order parameter to identify solid-like particles
($\bar{q}_{3}$) does not affect the observed response of the cluster to
temperature. This is shown in Fig.~\ref{q3op}, where the number of particles in
the cluster is monitored with two different order parameters for three
different temperatures.  Both order parameters allow to conclude that the
inserted cluster is critical between 235 and 240 K.
Obviously, the number of particles that belong to the cluster does depend on the order parameter. 
The order parameter we use in this work should at least work well for the inner particles 
given that, according to Fig.~1 of the main text, it is able to 
discriminate between bulk liquid and bulk solid particles. The main ambiguity in the number of particles
belonging to the cluster comes from those particles that lie in the interface. 
In view of Fig.~3 of the main text, it seems that our order parameter is doing reasonably well 
in identifying such particles either. Nonetheless, we have considered an error as large
as 60\% in the identification of the {\em interfacial particles} to estimate the error of the nucleation 
rate. In this way the unavoidable ambiguity in the determination of the cluster size is reflected
in the error bar of $J$.

\subsection{Cluster size and potential energy versus time for all system sizes and model potentials studied}

\begin{figure}[h!]
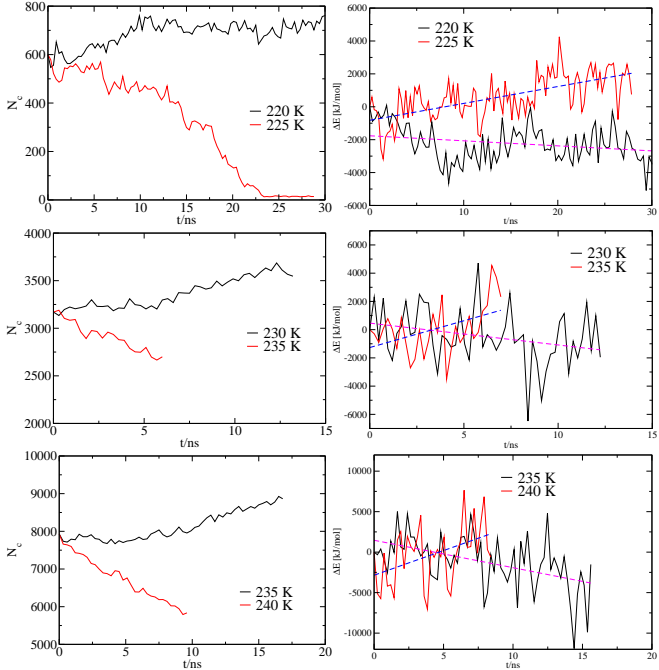

\includegraphics[clip,width=0.5\columnwidth]{22712-clusters-TIP4P2005.eps}%
\includegraphics[clip,width=0.5\columnwidth]{deltaE-661-tip4p2005.eps}
\includegraphics[clip,width=0.5\columnwidth]{76781-clusters-TIP4P2005.eps}%
\includegraphics[clip,width=0.5\columnwidth]{deltaE-4648-tip4p2005.eps}
\includegraphics[clip,width=0.5\columnwidth]{182585-clusters-TIP4P2005.eps}%
\includegraphics[clip,width=0.5\columnwidth]{deltaE-7805-tip4p2005.eps}
\caption{Left-hand panels: Number of molecules in the ice cluster versus time  for system B (top), L (middle) and H (bottom) simulated with the TIP4P/2005 potential. 
Right-hand panels: energy difference (between the energy at time $t$ and the one at time zero) versus time.  Results are shown for different temperatures as indicated in the legend.}
\label{fig:cluster-energy-2005}
\end{figure}

\begin{figure}[h!]
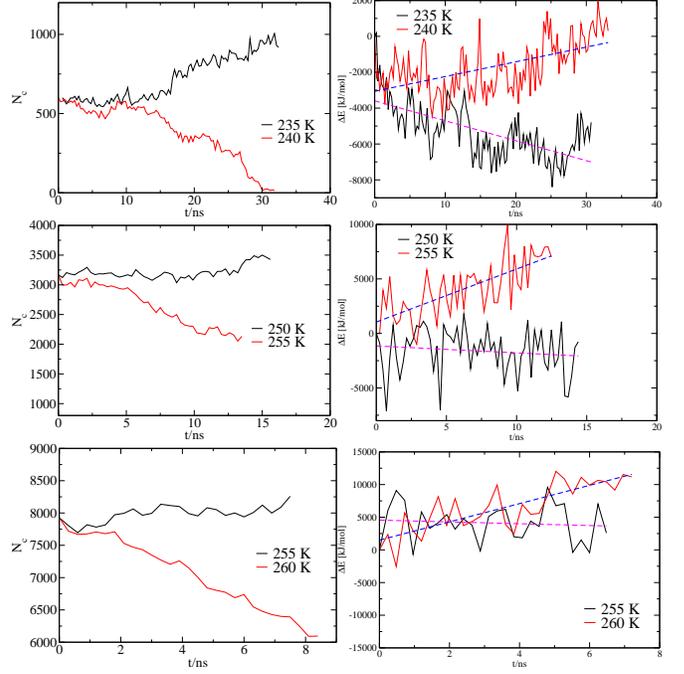

\includegraphics[clip,width=0.5\columnwidth]{22712-clusters-tip4pice.eps}%
\includegraphics[clip,width=0.5\columnwidth]{deltaE-661-tip4pice.eps}
\includegraphics[clip,width=0.5\columnwidth]{76781_clusters_tip4pice.eps}%
\includegraphics[clip,width=0.5\columnwidth]{deltaE-4648-tip4pice.eps}
\includegraphics[clip,width=0.5\columnwidth]{182585-clusters-tip4pice.eps}%
\includegraphics[clip,width=0.5\columnwidth]{deltaE-7805-tip4pice.eps}
\caption{Same as Fig. \ref{fig:cluster-energy-2005} but for TIP4P/Ice instead of TIP4P/2005.}
\label{fig:cluster-energy-ice}
\end{figure}

In order to determine the temperature at which the cluster was critical, we
evaluated the highest temperature at which the cluster grows and the lowest
temperature at which it melts. In Fig.~4 of the main text we represented the
number of molecules in the ice cluster versus time for system H simulated with
TIP4P/2005 potential. In what follows, we present the results of the cluster
size versus time for all sizes (B,L and H) for both the TIP4P/2005
(Fig.\ref{fig:cluster-energy-2005}) and TIP4P/Ice water models
(Fig.\ref{fig:cluster-energy-ice}).  An analogous result could have been
obtained by monitoring the potential energy of the system as a function of time
(see the right panels at Figs.~\ref{fig:cluster-energy-2005} and
\ref{fig:cluster-energy-ice}).

The energy is much less sensitive to changes in the cluster size than the order parameter. 
This is due to the fact that the number of molecules in the cluster is a small fraction of the total number of molecules.  
Nonetheless, by making a linear fit to the time evolution of the energy
we obtain in all cases consistent results with the analysis based in the order parameter: 
the cluster grows when the slope is negative and shrinks when it is positive. 

\section{Acknowledgments}
E. Sanz acknowledges financial support from the EU grant 322326-COSAAC-FP7-PEOPLE-2012-CIG 
and from the Spanish grant Ramon y Cajal. C. Valeriani acknowledges financial support
from the EU grant 303941-ANISOKINEQ-FP7-PEOPLE-2011-CIG and from the Spanish grant Juan de la Cierva. 
Fundings also come from MECD Project FIS2010-16159 and from CAM  MODELICO P2009/ESP/1691. 
All authors acknowledge the use of the super-computational facility Tirant at Valencia  
from the Spanish Supercomputing Network (RES), along with the technical support and the
generous allocation of CPU time to carry out this project (through projects QCM-2012-2-0017,
QCM-2012-3-0038 and QCM-2013-1-0047).
One of us (C. Vega) would like to dedicate this paper to the memory of Prof. Tomas Boublik.
We thank the three referees for their useful comments and to A. Reinhardt and J.P.K. Doye for sending us a preprint
of their work prior to publication. 

\end{document}